\documentclass[aps,twocolumn,floatfix, bibnotes]{revtex4}
\usepackage{graphicx}%
\usepackage{dcolumn}
\usepackage{color}

\usepackage{amsmath}
\begin{document}
\begin{sloppy}
\newcommand{\be}{\begin{equation}}
\newcommand{\ee}{\end{equation}}
\newcommand{\bea}{\begin{eqnarray}}
\newcommand{\eea}{\end{eqnarray}}
\newcommand\bibtranslation[1]{English translation: {#1}}
\newcommand\bibfollowup[1]{{#1}}

\newcommand\pictc[5]{\begin{figure}
                       \centerline{
                       \includegraphics[width=#1\columnwidth]{#3}}
                   \protect\caption{\protect\label{fig:#4} #5}
                    \end{figure}            }
\newcommand\pict[4][1.]{\pictc{#1}{!tb}{#2}{#3}{#4}}
\newcommand\rpict[1]{\ref{fig:#1}}

\newcommand\leqt[1]{\protect\label{eq:#1}}
\newcommand\reqtn[1]{\ref{eq:#1}}\newcommand\pictFig[1]{\pagebreak \centerline{
                   \includegraphics[width=\columnwidth]{#1}}
                   \vspace*{2cm}
                   \centerline{Fig. \protect\addtocounter{Fig}{1}\theFig.}}
\newcommand\reqt[1]{(\reqtn{#1})}

\newcounter{Fig}

\title{Ideal and nonideal electromagnetic cloaks}

\author{Nina A. Zharova$^{1,2}$, Ilya V. Shadrivov$^{1}$, Alexander A. Zharov$^{1,3}$, and Yuri S. Kivshar$^1$}
\affiliation{$^1$Nonlinear Physics Center, Research School of
Physical Sciences and Engineering,  Australian National University,
Canberra ACT 0200, Australia\\
$^2$Institute of Applied Physics, Russian Academy of Sciences, Nizhny Novgorod 603600, Russia \\
$^3$Institute for Physics of Microstructures, Russian Academy of
Sciences, Nizhny Novgorod 603950, Russia}

\begin{abstract}
We employ the analytical results for the spatial transformation of
the electromagnetic fields to obtain and analyze explicit
expressions for the structure of the electromagnetic fields in
invisibility cloaks, beam splitters, and field concentrators. We
study the efficiency of nonideal electromagnetic cloaks and discuss
the effect of scattering losses on the cloak invisibility.
\end{abstract}


\maketitle

Cloaking of objects was first explicitly suggested by
Pendry~\cite{1} as a method to cover an object by a layer of a
composite material with varying characteristics in such a way that
the scattering of electromagnetic waves from this object vanishes,
so that the structure of the electromagnetic field outside the cloak
appears as if the object and cloak are absent. The required material
for creating such invisibility cloaks should have rather complex
anisotropic, spatially varying properties. Moreover, such a material
should possess both nontrivial dielectric and magnetic responses.
Generally speaking, it is impossible to find dielectric media with
the required varying properties in nature, and therefore the first
electromagnetic cloak realized experimentally employed microwave
metamaterials~\cite{2}. Metamaterials are artificial microstructured
materials with electric and magnetic properties designed by careful
engineering of their constituents. The constituents of these
composite materials are normally sub-wavelength electric and
magnetic resonators, with the specific properties determined by
their geometry. The metamaterials can also be anisotropic, and thus
they can satisfy, in principle, all the requirements for realizing
invisibility cloaks for electromagnetic waves.

The concept of the electromagnetic cloaking is based on the
coordinate transformation of space and the corresponding
transformation of Maxwell's equations, which provide the required
expressions for the effective spatially varying dielectric
permittivity and magnetic permeability of the cloak medium~\cite{1}.
The possibility of creating such cloaks has been confirmed
experimentally~\cite{2}, and it was further discussed and
demonstrated in many numerical studies~\cite{3,4,5,6,14,15,18,20}.
Numerical simulations are usually performed either by using
commercial finite-element equation solvers~\cite{1,14,15,18,20} or
by employing the decomposition of the electromagnetic fields into a
set of the corresponding eigenmodes~\cite{3,4,5,6}. Several studies
have analyzed a possibility of creating simplified
cloaks~\cite{14,18}, since the required media parameters for an
ideal cloaking are extremely complex for their realization in
experiment. In particular, an optimization of the cloaking
parameters for suppressing scattering has been analyzed in
Refs.~\cite{5,18}. In this context, it is worth mentioning an
earlier pioneering study~\cite{19}, where an analogy of the
coordinate transformation to an effective magneto-dielectric medium
was discussed for some particular cases.

In this Letter we analyze the explicit analytical expressions for
the structure of the effective electromagnetic fields caused by the
spatially varying coordinate transformations. In a sharp contrast to
the rather formal transformations of Maxwell's equations in the
Minkovskii form~\cite{32} used in the original pioneering
paper~\cite{1} and all subsequent publications, we present a simpler
form of the time-independent transformations that can be immediately
employed for solving many important problems. Based on these results
we suggest a simple recipe for evaluating the screening efficiency
of non-ideal cloaks and calculate their scattering dipole moment.

{\em General formalism.} We start our analysis by considering the
propagation of monochromatic (i.e. $\sim \exp (i \omega t)$)
electromagnetic fields described by Maxwell's equations
\begin{eqnarray}
\label{eq1}
{\rm curl} \, {\bf H} = \frac{i \omega}{c}{\bf D} \;\; ,
{\rm curl} \, {\bf E} =-\frac{i \omega}{c}{\bf B} \ .
\end{eqnarray}
First, Eq.~(\ref{eq1}) can be written in the component notations in
the following form,
\begin{equation}
E^{abc} {\bf x}_a \frac{\partial}{\partial x^b} H_c = \frac{i \omega}{c}{\bf x}_a D^a \ ,
\label{Ampere}
\end{equation}
where $E^{abc}$ is the totally antisymmetric symbol~\cite{footnote},
${\bf x}_a$ is the orthonormal basis corresponding to the Cartesian
coordinates, the indices $(a, b, c)$ run through the values $(1, 2,
3)$, and $x^a$ are the Cartesian coordinates corresponding to the
orthonormal basis ${\bf x}_a$. In a medium with the dielectric
permittivity $\hat\varepsilon_0$, the electric displacement vector
$\bf D$ can be expressed in terms of the electric field as ${D^a} =
\hat\varepsilon^{ab}_0{ E_b}$. The magnetic displacement vector $\bf
B$ can be expressed as ${B^a} = \hat\mu^{ab}_0{ H_b}$, in a medium
with magnetic permeability $\hat\mu_0$. We note that the repeated
upper and lower indices imply summation (Einstein convention).

Now we introduce new coordinates $\bf y$ through the transformations
${\bf y}={\bf y}({\bf x})$ with the metric tensor $g^{ij}$ defined
in a standard way,
\begin{equation}
\label{metric}
g^{ij}(x) = \frac{\partial y^i }{\partial x^a}\frac{\partial y^j}{\partial x^b }\delta^{ab} \ ,
\end{equation}
where $\delta^{ab}$ is Kronecker's  delta-function. The determinant
$g(x)=\det||g^{ij}(x)||$ is of a particular importance for the
coordinate transformations.

In the framework of these new coordinates the l.h.s. of
Eq.~(\ref{Ampere}) can be re-written in the form,
\begin{eqnarray}
\label{002} \frac{1}{2}E^{abc}\left (\frac{\partial y^i}{\partial
x^b}\frac{\partial H_c}{\partial y^i} -\frac{\partial y^j}{\partial
x^c}\frac{\partial H_b}{\partial y^j} \right )
= \nonumber\\
\frac{1}{2}E^{abc}\left (\frac{\partial y^i}{\partial
x^b}\frac{\partial y^k}{\partial x^c} \frac{\partial \tilde
H_k}{\partial y^i} -\frac{\partial y^k}{\partial x^c}\frac{\partial
y^i}{\partial x^b} \frac{\partial \tilde H_i}{\partial y^k} \right
)\ ,
\end{eqnarray}
where $\bf \tilde H$ is the transformed magnetic field defined as
\begin{equation}
\label{main} {\tilde H}_k=\frac{\partial x^c}{\partial y^k}H_c,
\;\;\; H_c=\frac{\partial y^k}{\partial x^c}{\tilde H}_k.
\end{equation}
The totally antisymmetric symbol is a tensor density, and real
tensor (Levi-Civita tensor, $\epsilon^{ijk}$), which coincides with
$E^{ijk}$ in Cartesian coordinates, is related to the $E^{ijk}$
through $\epsilon^{ijk}=\sqrt{g}E^{ijk}$, and it transforms under
coordinate transformation according to the tensor rules
\begin{equation}
\label{dens}
\epsilon^{lik} =
\frac{\partial y^l}{\partial x^a}\frac{\partial y^i}{\partial x^b}
\frac{\partial y^k}{\partial x^c}E^{abc} =\sqrt{g}E^{lik} \ .
\end{equation}

Multiplying both parts of Eq.~(\ref{Ampere}) by $\partial
y^l/\partial x^a$ and using Eq.~(\ref{dens}), we obtain
$$
\frac{1}{2}\sqrt{g}
\left [
E^{lik}\frac{\partial \tilde H_k}{\partial y^i}
-E^{lki}\frac{\partial \tilde H_i}{\partial y^k}\right ]
=\frac{i \omega}{c}\frac{\partial y^l}{\partial x^a}\varepsilon_0^{ab}
\frac{\partial y^n}{\partial x^b}{\tilde E}_n \ ,
$$
where $\bf \tilde E$ is the transformed electric field defined as
\begin{equation}
\label{main1}
{\tilde E}_n=\frac{\partial x^b}{\partial y^n}E_b \  ,  \
E_b=\frac{\partial y^n}{\partial x^b}{\tilde E}_n \ .
\end{equation}
As a result, we arrive at the transformed equation
\begin{equation}
E^{lik}\frac{\partial \tilde H_k}{\partial y^i} =({\rm curl}{\bf
\tilde H})^l =\frac{i \omega}{c}\frac{1}{\sqrt{g}} \frac{\partial
y^l}{\partial x^a}\frac{\partial y^n} {\partial
x^b}\varepsilon_0^{ab}{\tilde E}_n \ , \label{final_eq}
\end{equation}
which coincides with the original Maxwell's equations but it is
written for the new electromagnetic fields in the new (spatially
transformed) space. The left-hand side of this equation contains the
curl operation on the transformed magnetic field, whereas in the
right-hand side of Eq.~(\ref{final_eq}) we have a new electric
displacement vector which is defined as the product of the new
dielectric tensor,
\begin{equation}
\label{eq_ep}
 \varepsilon^{ ln}_{\rm
eff}=\frac{1}{\sqrt{g}}\frac{\partial y^l}{\partial
x^a}\frac{\partial y^n} {\partial x^b}\varepsilon_0^{ab} \ ,
\end{equation}
and the new electric field~(\ref{main1}).
Applying the same transformation procedure to the second equation of
Eq.~(\ref{eq1}), we obtain the corresponding
expression for the effective magnetic permeability,
\begin{equation}
\label{eq_mu}
\mu^{ ln}_{\rm eff}=\frac{1}{\sqrt{g}}\frac{\partial
y^l}{\partial x^a}\frac{\partial y^n} {\partial x^b}\mu_0^{ab}.
\end{equation}
When the initial medium represents vacuum
($\varepsilon_0^{ab}=\mu_0^{ab}=\delta^{ab}$), the resulting new
dielectric tensor coincides with the effective magnetic tensor.


Our results for the effective material parameters coincide with
those of Ref.~\cite{13} but, in addition, they also provide {\em
explicit analytical formulas} for the transformed fields. Thus, we
can find analytically the field distribution for any specific cloak
{\em without} solving Maxwell's equations in the transformed
geometry, but just transforming the free-space fields. We have both
the expressions for the structure of the cloak and equations
(\ref{main}) and (\ref{main1}) for the field transformations in the
cloak.  As a straightforward application, our results allow
analytical estimations of the scattering from non-ideal cloaks, and
thus a potential optimization of their parameters.

{\em Examples.} We use those results for calculating the field
structure in both cylindrical and spherical cloaks, field
concentrators, and other devices designed via the transformation
optics. As the first example, we consider the linear coordinate
transformation for the cylindrical cloak~\cite{13}, $X=X(x,y)$,
$Y=Y(x,y)$, $Z=z$, where the radius is transformed as $R=a +
r(b-a)/b$, where $R=\sqrt{X^2+Y^2}$, $r=\sqrt{x^2+y^2}$.  The cloak
occupies the space $a<R<b$, where $a$ and $b$ are the inner and
outer radii of the cloak, respectively. To find the field
distribution in the cloak illuminated by a plane electromagnetic
wave, we write the plane wave in vacuum as $H_x=\exp (i \kappa_y
y)$, $H_y=0$,  $E_z=\exp (i \kappa_y y)$. Applying now the
transformations according to Eqs.~(\ref{main}), we find the field
distribution in the electromagnetic cloak with the material
parameters (\ref{eq_ep}) and (\ref{eq_mu}), $\tilde  E_z=\exp (i
\kappa_y Y)[b \left(b-a \right)^{-1} ( 1- a R^{-1})]$, $ \tilde H_x
= b( b-a)^{-1}( 1-aY^2 R^{-3}) \tilde E_z$, $\tilde H_y = abXY
(b-a)^{-1} R^{-3} \tilde E_z$, for $a<R<b$. The field vanishes
inside the cloaked area, whereas outside of the cloak the field is
an unperturbed plane wave.
At the external surface of the cloak (at $R=b$), the normal
component of the magnetic field is discontinuous, since the linear space
transformation produces discontinuities of the cloak material
parameters. Using the transformation function
\begin{equation}
r=b-a\left (\frac{b-R}{b-a}\right )^\beta
\label{smooth}
\end{equation}
provides, for large enough values of $\beta$ ($\beta > 1$)
continuity of the field components at the external
surface of the cloak.
\begin{figure}
\includegraphics[width=\columnwidth]{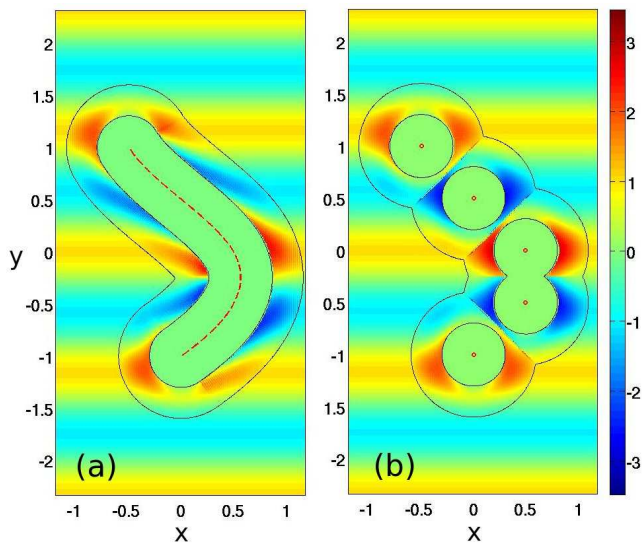}
\caption{\label{fig1}(color online) Structure of the magnetic field
$H_x$ for (a) an arbitrary-shaped cloak and (b) multiple cloaks
produced by the space transformation in accord with
Eq.~(\ref{smooth}), where we take the parameters $a=0.3$, $b=0.6$,
and $\beta =2$. The cloak is illuminated by a plane wave,
$H_x^0=\exp (i \kappa_y y)$ for $\kappa_y=5.4$. Center of the cloak
is shown by red line, in the case (a), and dots, in the case (b). }
\end{figure}
As a matter of fact, Eqs.~(\ref{main}) and (\ref{main1}) allow us to
find the field structure for {\em an arbitrary continuous space
transformation}. Though for an arbitrary transformation one cannot
find the analytical expressions for the electromagnetic fields,
simple numerical calculations provide the required results without
solving Maxwell's equations.

As an example of a cloak of complex shape, we consider cut-type
space transformation where the cloak covers an arbitrary area `cut'
from the space (see, e.g., Fig.~1). We define the internal surface
of the cloak as a set of all points which are closer than $b$ to the
cut line ${X_{\rm cut} (\xi ), Y_{\rm cut}(\xi )}$. For each point
$(X,Y)$ we find the corresponding closest point on the cut line
$(X^0_{\rm cut},Y^0_{\rm cut}) (X,Y)$ by finding the minima of the
function $[(X-X_{\rm cut}(\xi ))^2+(Y-Y_{\rm cut}(\xi ))^2]^{1/2}$
for all possible
 $\xi$. We define the distance from the 'cut' as
$R(X,Y)=[(X-X^0_{\rm cut})^2+(Y-Y^0_{\rm cut})^2]^{1/2}$ and create
the cloaking area $a<R<b$ by transforming the space according to
Eq.~(\ref{smooth}). In the area of the cloak, $a<R<b$, we define the
new coordinates $(x,y)$ as
$$
 x=X^0_{cut}+(X-X^0_{\rm cut})\frac{r}{R} \ ,  \  y=Y^0_{cut}+(Y-Y^0_{\rm cut})\frac{r}{R} \ ,
$$
which provide a conformal mapping of $(x,y)$ into $(X,Y)$.

Figure~1(a) shows the distribution of the $x$-component of the
magnetic field for the plane wave incident on the cloak. The cut is
created through the graphic input in Matlab, where we specify
several points (five points, in this example), which are connected
by a smooth line through the cubic spline function. Figure~1(b)
shows the field distribution in the cloaking problem with several
cylindrical cloaks centered at the same five points.

Parameters of transformation $r(R)$ are  $a=0.3$, $b=0.6$, $\beta
=2$. The field distribution is calculated for the incident plane
wave $H_x=\exp (i \kappa_y y)$ with $\kappa_y=5.4$.

We note that, despite the divergence of the dielectric and magnetic
functions at the internal interface of the cloak, the electric and
magnetic fields remain finite. For example, the radial and
tangential components of the magnetic field can be found as
$$
|H_R|=\frac{b}{(b-a)}\frac{|X|}{R} \ , \ |H_\tau
|=\frac{b}{(b-a)}\frac{|Y|}{R}\left (1-\frac{a}{R} \right ) .
$$
We note that $|H_R|$ is finite at the internal interface of the
cloak, while $|H_\tau |$ vanishes when $R \to a$.

From the uniqueness theorem of electromagnetism we can conclude that
since the tangential components of the electric or magnetic field at
the closed surface vanish, the field in the volume surrounded by
this surface vanishes as well; this means that we indeed have the
volume ($R < a$) concealed from the electromagnetic radiation.

{\em Effectiveness of non-ideal cloaks.} Above we demonstrated that
a cloak with the parameters defined by Eqs.~(\ref{eq_ep}) and
(\ref{eq_mu}) provides complete invisibility of the objects hidden
inside. Moreover, such a cloak works for an arbitrary transformation
function ${\bf y}={\bf y}({\bf x})$, either smooth or piece-wise
continuous~\cite{14}. However, this kind of ideal cloaking seems to
be impossible in practice, since it requires infinite values of
material parameters at the internal surface. The simplest way to
overcome this difficulty is to truncate the transformed area. For
example, for the coordinate transformation with the internal radius
$a$ we create the cloak only for $R > a+\delta a$, and then place a
perfectly conducting metal surface at $R = a+\delta a$. Now, if we
make an inverse transformation from the cloak to vacuum, the
metallic surface transforms into a metallic cylinder with the radius
$r_{\rm eff}$. If $dR/dr|_{r=a} \ne 0$, then $r_{\rm
eff}\approx\delta a (dR/dr|_{r=a})^{-1}$, {\em the simplified cloak
scatters light} as a metallic cylinder of the radius $r_{\rm eff}$,
which can be much smaller than the size of the hidden region $a$.

One of the main limitations of the resonant medium used for creating
electromagnetic cloaks is losses. To estimate the effect of small
losses on the cloak performance, we assume that $\hat \varepsilon =
\hat \mu =\hat \varepsilon^{\prime} + i\hat \varepsilon^{\prime
\prime}$, $\hat\varepsilon^{\prime \prime} \ll \hat
\varepsilon^{\prime}$, whereas $\hat\varepsilon^{\prime}$ is given
by Eq.~(\ref{eq_ep}). We consider a plane wave incident on such a
cloak. After applying the inverse coordinate transformation, which
transforms the electromagnetic fields into vacuum plane waves, we
obtain in original space the excited effective electric and magnetic
currents dependent on vacuum electric and magnetic fields,
$$
j_{e(m)}^a=\sigma^{ab} E_b^v(H_b^v), \;\;
\sigma^{ab}=-\frac{\omega\sqrt{g}}{4\pi} \frac{\partial
x^a}{\partial y^i}\frac{\partial x^b}{\partial y^j }
{\varepsilon^{ij}}^{\prime \prime}.
$$
These currents lead to the scattering radiation which can be easily
evaluated either analytically or numerically.

Next, we can estimate the effect of mismatched dielectric
permittivity and magnetic permeability on the cloak performance. We
assume that $\hat \varepsilon =\hat \mu+ \delta \hat
\varepsilon=\hat \varepsilon_0+\delta \hat \varepsilon$, where $\hat
\varepsilon_0$ corresponds to an ideal cloak and $|\delta \hat
\varepsilon| \ll |\hat \varepsilon_0|$. Then, we apply the
inverse transformation ${\bf x} \to {\bf y}$ and find that in the
original space the dielectric permittivity differs from the unity by the
value
$$
\delta \tilde \varepsilon^{ab}=\sqrt{g} \frac{\partial x^a}{\partial
y^i}\frac{\partial x^b}{\partial y^j } \delta \varepsilon^{ij}. $$
If we assume that $(\hat \varepsilon - \hat \mu) \propto \hat
\varepsilon$, namely $\delta  \varepsilon^{ij}=\alpha
\varepsilon^{ij}_{0}$ (for $\alpha \ll 1$),  then $\delta \tilde
\varepsilon^{ab}=\alpha\delta^{ab}$.  Thus, in this case the
non-ideal cloak scatters the waves as an object (cylinder of the
radius $b$ for the cylindrical cloak) with a scalar dielectric
permittivity $1+\alpha \approx 1$.

Now we consider $\delta  \varepsilon^{ij}=\alpha \delta^{ij}$, for
$\alpha \ll 1$.  This gives us a relatively simple result for
$\delta \tilde \varepsilon^{ab}$,
\begin{equation}
\delta \tilde \varepsilon^{ab}=\alpha\sqrt{g}
\frac{\partial x^a}{\partial y^i}\frac{\partial x^b}{\partial y^i }
\end{equation}
and also an expression for the electric current
\begin{equation}
\label{eff_curr}
{\bf j}_{\rm eff}\approx \frac{i\omega}{4\pi}\alpha\hat
\varepsilon^{-1} {\bf E}^v,
\end{equation}
which can be used  for calculating the scattered fields. It is worth
mentioning that in this case the cloak is sensitive to small
perturbations of the parameters. Indeed, the inverse dielectric
permittivity has a singularity at the inner surface of the cloak, so
if the truncation value $\delta a$ is small then the parameter
$\alpha$ has to be very small as well so that it does not perturb
the structure of the plane wave. This result has been predicted
earlier in Ref.~\cite{4}. We also note that very similar
expressions for the electric and magnetic currents in the lossy case
may also lead (for particular forms of $\hat \varepsilon^{\prime
\prime}$) to such a strong sensitivity.

As an example let us consider plane wave scattering on a slightly
non-ideal cloak with the magnetic permeability  tensor components
$\mu_{rr}=(R-a)/R$, $\mu_{\phi\phi}=R/(R-a)$,
$\mu_{zz}=\gamma^{-2}(R-a)/R$, $\gamma=(b-a)/b$ and dielectric
permittivity tensor $\varepsilon_{ii}=\mu_{ii}+ \alpha $, $\alpha
\ll 1$. We suppose that the wave propagates in $y$-direction, its
wavenumber, $k_y =5.4$, and inner and outer radii of the cloak are
$a=0.3$, $b=0.6$, respectively. For the case of $\alpha =0$ such
cloak is ideal, and it corresponds to the linear coordinate
transformation $R = a+ \gamma r$, $0 < r < b$. In order to avoid the
singularity we place a metallic cylinder at $R = a+\delta a$, which
conceals the inner region of the cloak but it also distorts
scattering performance. Scattered fields are regarded as weak, and
we use the perturbation approach to calculate them. Let us make an
inverse coordinate transformation $R \to r$, so that the metal
screen radius transforms to $r_0=\delta a /\gamma$. We calculate the
effective current according to the Eq.~(\ref{eff_curr}) and find the
scattering field pattern for both $TE$ (${\bf E}^0 = {\bf z}^0 E$)
and $TM$ (${\bf H}^0 = {\bf z}^0 H$) waves (see Figures 2a,b).
\begin{figure}
\includegraphics[width=0.95\columnwidth]{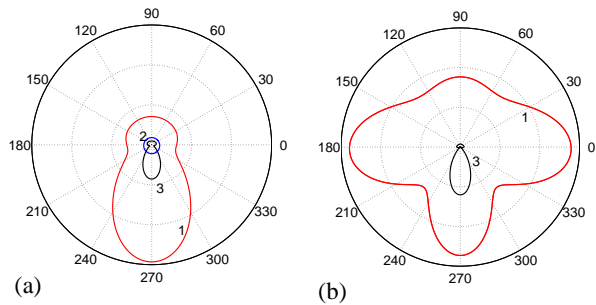}
\caption{\label{fig2}(Color online) Scattering patterns for
non-ideal cloak for $TE$ (a) and $TM$ (b) waves. Red lines (1)
indicate scattering from bare metallic cylinder, while blue line (2)
corresponds to the scattering from the cloaked cylinder with truncation level $r_0=0.05 b$ and $\alpha =0$. Black curves (3) correspond to the scattering (mainly forward) on the
dielectric cloak with parameter $\alpha = 0.1$ for the figure (a) and $\alpha =0.02$ for the figure (b). }
\end{figure}
One can see that for the $TM$ polarization the scattering from
non-ideal cloak is negligible, and in the shown scale it is
represented by a point in the center of the Fig.~(\ref{fig2})(b). If
the truncation level, $r_0$,  is reduced, the scattered radiation
for $TM$ wave is also reduced. On the contrary, $TE$ wave scattering
is significantly enhanced, e.g., if we decrease $r_0$ to $0.005 b$,
than the scattered power becomes four times larger.

This approach allows to estimate the scattering losses in an
arbitrary nonideal cloak (e.g., such as that shown in Fig.~1), since
the effective currents can be calculated using Eq.~(\ref{eff_curr})
with the plane wave excitation.

In conclusion, we have employed the analytical results for the
distribution of electromagnetic fields in the transformed
coordinates to analyze an explicit solution for the problem of the
wave scattering in electromagnetic cloaks, beam splitters, energy
concentrators, and other devices of the transformation optics.
Employing those results, we have derived a simple criterion of the
efficiency of nonideal cloaks based on estimations of wave
scattering.

The authors thank Prof. V.E. Semenov for valuable discussions. This
work was supported by the Australian Research Council and the
Russian Fund for Basic Research (grant No. 08-02-00379).

\end{sloppy}

\begin{thebibliography}{99}
\bibitem{footnote}
We remind that the components $E^{abc}$ differ from
zero only for the indices  $a\ne b\ne c$; the index permutation
changes the sign of the totally antisymmetric symbol, and $E^{123}=1$.
\end{thebibliography}
\end{document}